\documentclass[final,5p,times]{elsarticle}
\usepackage{graphicx}
\usepackage{float}
\usepackage{amsmath}
\usepackage{url}
\usepackage{lineno,hyperref}

\journal{Nuclear Instruments and Methods A}

\begin{document}
	\begin{frontmatter}
	\title{Electron transport and electric field simulations in two-phase detectors with THGEM electrodes}

	\author[BINP,NSU]{A. Bondar}
	\author[BINP,NSU]{A. Buzulutskov}
	\author[BINP,NSU]{E. Frolov\corref{cor1}}
	\ead{geffdroid@gmail.com}
	\author[BINP,NSU]{V. Oleynikov}
	\author[BINP,NSU]{E. Shemyakina}
	\author[BINP,NSU]{A. Sokolov}
		
	\cortext[cor1]{Corresponding author}
	\address[BINP]{Budker Institute of Nuclear Physics SB RAS, Lavrentiev avenue 11, 630090 Novosibirsk, Russia}
	\address[NSU]{Novosibirsk State University, Pirogova st. 2, Novosibirsk 630090, Russia}
		\begin{abstract}
		One of the main features of two-phase detectors with electroluminescence (EL) gap being developed in our laboratory for dark matter search is the extensive use of THGEMs (Thick Gas Electron Multipliers). In various versions of the detector, the THGEMs are used as electrodes in the gas and liquid phases to form the drift, electron emission and  EL regions, as well as  for avalanche amplification of a signal in the gas phase. In this work the simulations of the electric field and electron transport through such THGEM electrodes were performed. In the liquid phase,  these simulations allowed to determine the optimal parameters, such as the hole diameter of THGEM and applied voltage across it, that can provide effective transmission of the electrons from the drift region to that of the EL gap. In the gas phase, the effect of the THGHEM voltage on the electric field uniformity in the EL gap was studied. 
		\end{abstract}

	\end{frontmatter}  


  \section{Introduction}
Two-phase detectors with electroluminescence (EL) gap, based on Ar and Xe, are relevant to experiments for direct search of dark matter particles \cite{Chepel:survey,DarkSide20k18}. Several versions of such detectors were developed in our laboratory for the study of EL mechanism in pure argon \cite{NBrS18} and argon doped with nitrogen \cite{ArN2CRAD17,ELReview17}, for SiPM-matrix readout of two-phase detectors, directly \cite{SiPMMatrix19} or using combined THGEM/SiPM-matrix multipliers \cite{SiPMMatrix19,CRADRev12,SiPMMatrix13}, and for the measurements of ionization yields in liquid argon \cite{IonYield17}. One of the main features of two-phase detectors developed in our laboratory is the extensive use of THGEMs (Thick Gas Electron Multipliers \cite{BreskinConciseRev}). The THGEMs are used in the liquid and gas phases to form the drift, electron emission and electroluminescence regions, as well as for avalanche amplification of a signal in the gas phase. 

In this work the simulations of electron transport and electric fields in the two-phase detector were conducted to determine the optimal THGEM parameters for effective transmission of the electrons from the liquid phase into the EL gap and their transport to the anode of the two-phase detector. Furthermore, the simulations allowed us to ascertain the degree and extent of field uniformity in those field regions.

  \begin{figure}[t!]
	\includegraphics[width=\linewidth]{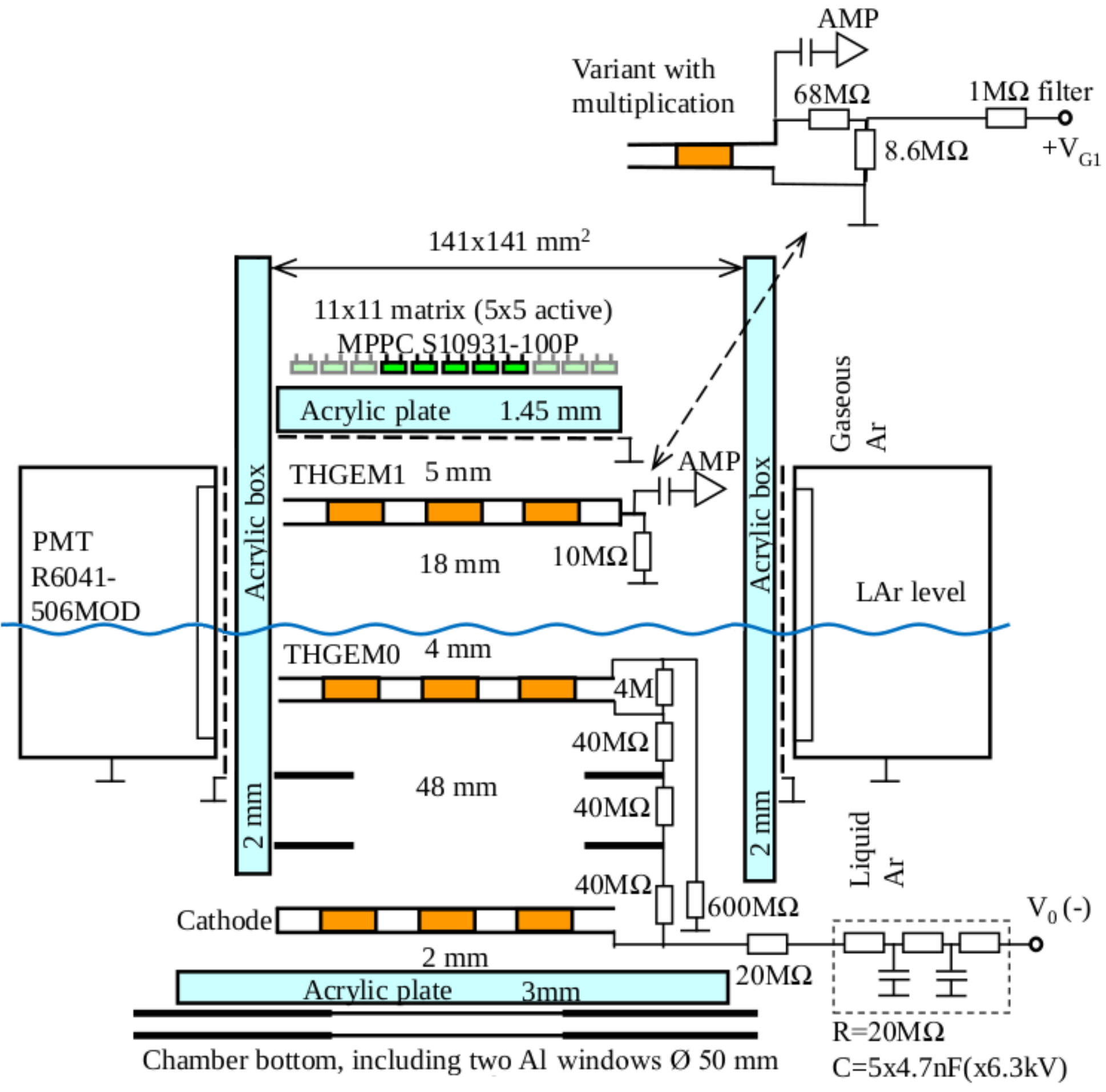}
	\caption{Generalized version of the two-phase detector with EL gap used in \cite{NBrS18,ArN2CRAD17,SiPMMatrix19,IonYield17} (not to scale). The 4 M\(\Omega\) resistance defining the voltage across THGEM0 is referred to as R$_{THGEM0}$. For THGEM1, two readout configurations are shown in the figure, reflecting those when THGEM1 acted as either an anode of the EL gap, or electron multiplication element of the THGEM/SiPM-matrix multiplier. 
	} 
	\vspace{-10pt}
	\label{fig:setup_scheme}
  \end{figure}
 
It is worth mentioning previous works on simulation of the electric fields and electron transport in THGEM-like structures in the gas phase: \cite{Shalem2006, Cantini2015, Bhattacharya2017, Correia2018}. To the best of our knowledge, no one has simulated THGEM performance in liquid argon thus far; it has been done for the first time in the present work.

  \section{Structure of the THGEM-based two-phase detector}
 
The generalized version of the two-phase detector with EL gap used in our current studies \cite{NBrS18,ArN2CRAD17,SiPMMatrix19,IonYield17} is shown in Fig. \ref{fig:setup_scheme}. To form the drift, electron emission and electroluminescence regions, THGEMs were used instead of more conventional wire grids, providing the advantage of electrode rigidity  that allowed to avoid the problem of wire grid sagging \cite{PatentCRAD}. The detector included a cathode electrode, field-shaping electrodes and THGEM0 (interface THGEM), immersed in a 55 mm thick liquid Ar layer. These elements were biased through a resistive high-voltage divider placed within the liquid, forming a 48 mm long drift region in liquid Ar. A 4 mm thick liquid Ar layer above the THGEM0 acted as an electron emission region. 

THGEM1 shown in Fig. \ref{fig:setup_scheme} was placed in the gas phase above the liquid and acted either as an anode of the EL gap (grounded through a resistor) or an electron multiplication element of the combined THGEM/SiPM-matrix multiplier (i.e. operated in electron avalanche mode) coupled to the EL gap. All electrodes had the same active area of 10$\times$10 cm$^2$. In the present detector version the THGEM geometrical parameters were similar to those of \cite{THGEMPaper13}: dielectric thickness of 0.4 mm, hole pitch of 0.9 mm, hole diameter of 0.5 mm and hole rim of 0.1 mm.

The voltage applied to the divider may vary from 3 to 22 kV providing the electric field of 0.093-0.68 kV/cm in liquid Ar in the drift region (\(E_{drift}\), between the cathode and THGEM0), 0.71-5.2 kV/cm in liquid Ar in the electron emission region (\(E_{emission}\), above THGEM0) and 1.1-8.0 kV/cm in gaseous Ar in the EL region (\(E_{EL}\), between the liquid surface and THGEM1). 

Using such a voltage divider, the THGEM0 was biased in a way to provide the transmission of drifting electrons from the drift region to that of electron emission: the electrons drifted successively from a lower to higher electric field region. The bias voltage ($V_{THGEM0}$) is defined by the appropriate resistor of the divider ($R_{THGEM0}$). Further we use $E_{THGEM0}$ = $V_{THGEM0}$/\hspace{0pt}(THGEM0 dielectric thickness) as a figure of merit. This field is always larger the the real field in THGEM hole and they are closer to each other the closer hole diameter is to 0. For $R_{THGEM0}$ = 4 M$\Omega$ the ratio $E_{drift}$:\hspace{0pt}$E_{THGEM0}$:\hspace{0pt}$E_{emission}$:\hspace{0pt}$E_{EL}$ is 1.0:\hspace{0pt}4.0:\hspace{0pt}7.6:\hspace{0pt}11.8.

Accordingly, the first objective of the present study was to determine the electron transmission through THGEM0 and its dependence on the electric field. The second objective was to determine the optimal parameters of THGEM0, such as the applied voltage and hole diameter, to increase the  electron transmission. The third objective was to study the effect of the THGEM0 and THGEM1 voltages on the field uniformity within the EL gap. The final objective was to determine field nonuniformity in the drift and EL regions caused by limited active area of the electrodes.

  \section{Simulation tools and parameters} \label{ToolsAndParameters}

In this work, the open-source tools Gmsh v3.0 \cite{Geuzaine2009}, Elmer v8.3 \cite{Ban2017, Elmer} and Garfield++ \cite{GarfieldPP} were used to respectively construct model mesh, calculate electric fields with FEM (Finite Element Method) and simulate electron drifting using drift velocity and diffusion coefficients of electrons in liquid Ar.

The experimental data for transverse and longitudinal electron diffusion coefficients are rather sparse and contradictory \cite{Derenzo:LAreDiffusion, Shibamura:LAreDiffusion, Cennini1994:LArTPC, Li2016:LAreDiffusion, Agnes2018} and this contributes the most to the uncertainty of electron transmission values. The measurements of transverse diffusion \cite{Shibamura:LAreDiffusion} were conducted only for the electric fields in 2-10 kV/cm range and that of longitudinal diffusion \cite{Li2016:LAreDiffusion} in the range of 0.1-1.5 kV/cm. Since the electric fields used in our simulation cover 0.093-5.2 kV/cm range, the extrapolation of data was necessary. The generalized Einstein relations \cite{Robson1972} were applied to calculate transverse diffusion coefficients at low fields using longitudinal ones, and vice versa at high fields:
	\begin{linenomath*}
	\begin{equation*}
	\frac{D_L}{D_T} = 1 + \frac{E}{\mu}\frac{\partial \mu}{\partial E}
	\end{equation*}
	\end{linenomath*}
where E is the electric field and $\mu$ is electron mobility directly related to drift velocity $\upsilon$ by
	\begin{linenomath*}	
	\begin{equation*}
	\upsilon = \mu E
	\end{equation*}
	\end{linenomath*}
It should be mentioned that the data on longitudinal diffusion at 100, 150 and 200 V/cm reported by DarkSide-50 \cite{Agnes2018} is significantly lower than those in \cite{Cennini1994:LArTPC} and \cite{Li2016:LAreDiffusion} while being closer to the values at zero electric field. Since the electron transmission is expected to decrease and become more dependent on $V_{0}$ with increased diffusion, we chose higher values of \cite{Cennini1994:LArTPC} and \cite{Li2016:LAreDiffusion} for our simulation to account for the worst-case scenario.

In contrast to the diffusion coefficients, electron drift velocity in liquid argon has been studied much better \cite{Cennini1994:LArTPC, Li2016:LAreDiffusion, Walkowiak:LAreVelocity, Sauli:LAreVelocity}. The global fit of electron mobility reported in \cite{Li2016:LAreDiffusion} was used for our simulations up to 2 kV/cm and the data of \cite{Walkowiak:LAreVelocity} above. For the details on diffusion coefficients and mobility of electrons in liquid argon the reader is referred to \cite{Li2016:LAreDiffusion}.

Also necessary for the simulations are dielectric constants of involved materials at 87 K. The $\varepsilon$ = 1.55 was used for liquid argon \cite{Aprile2006, Bolozdynya2010}, $\varepsilon$ = 4.4 for THGEM dielectric (glass-reinforced epoxy) \cite{Gerhold1998}, and $\varepsilon$ = 3.7 for acrylic (PMMA, Table 6.1 in \cite{Brydson1999}). The temperature dependence of acrylic's dielectric constant is neglected as other plastics' permittivity almost does not change with the temperature (Fig. 29 in \cite{Gerhold1998}).

\section{Simulation models}

\begin{figure}[t!]
	\includegraphics[width=\linewidth]{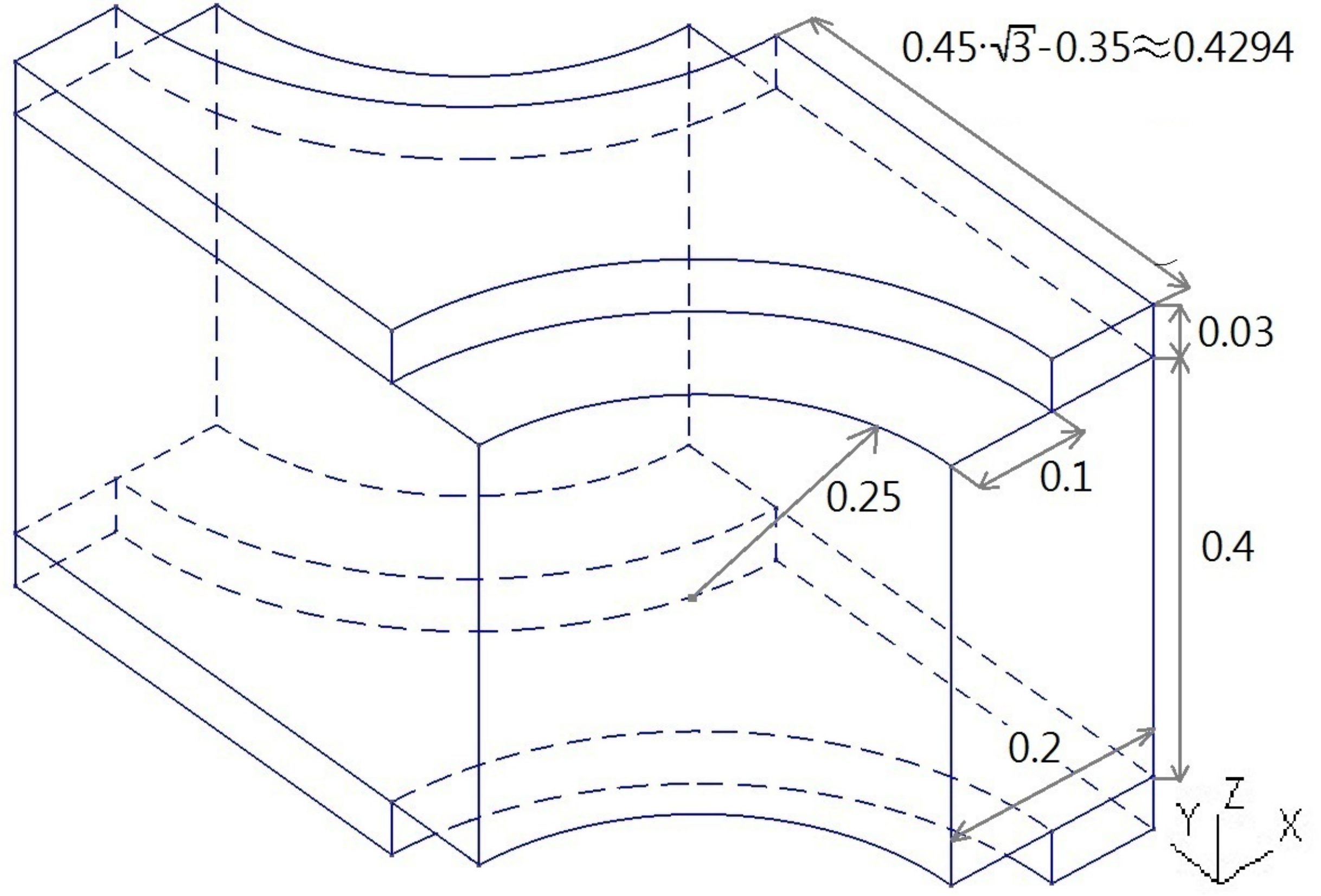}
	\caption{The model of 28\% THGEM cell used in calculations of electric fields (see Table~\ref{tab:THGEM_pars}). The cathode and anode planes are not depicted, all sizes are given in mm.}
	\label{fig:THGEM_cell_model}
	\end{figure}

THGEMs are generally much larger than their hole pitch which precludes simulation of THGEM as a whole due to tremendous number of elements (tetrahedrons) required for approximation of its holes with any reasonable detail. For this reason the periodicity of THGEM is utilized and only its small region (cell) shown in Fig. \ref{fig:THGEM_cell_model} is simulated. Applying antiperiodic boundary conditions allows to calculate electric fields in the infinite THGEM approximation. The distances from the THGEM to the cathode and anode planes are chosen to be large enough for the electric field to become uniform near them. Electron transmission through THGEM0 is defined as:
	\begin{linenomath*}
	\begin{equation*}
	T_e = \frac{\textrm{number of electrons that reached anode plane}}{\textrm{number of electrons generated in drift region}}
	\end{equation*}
	\end{linenomath*}

Note that electron emission probability at liquid-gas interface is not taken into consideration in this simulation because in two-phase Ar it was shown to be close to unity at emission fields exceeding 0.7 kV/cm \cite{Bondar2009:e_emission}.

\begin{figure}[t!]
	\vspace{-10pt}
		\begin{tabular}{c}
		\includegraphics[width=\linewidth]{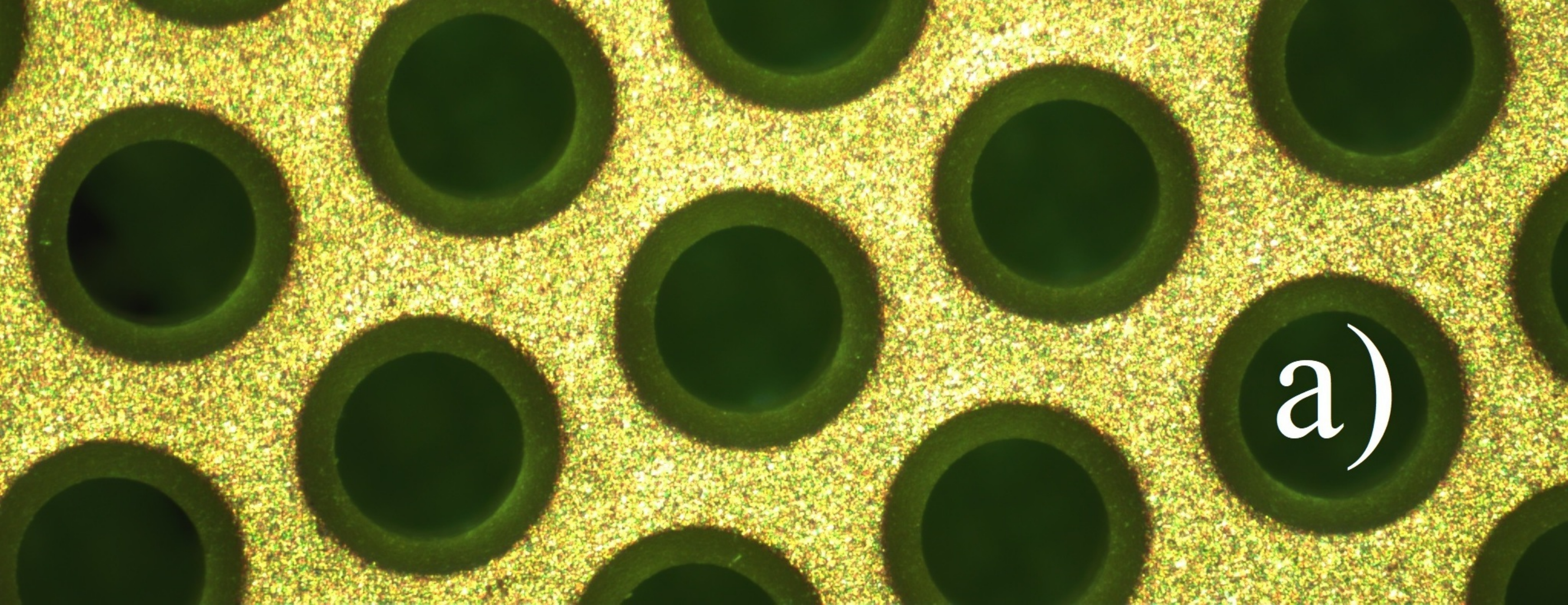}\\[\tabcolsep]
		\includegraphics[width=\linewidth]{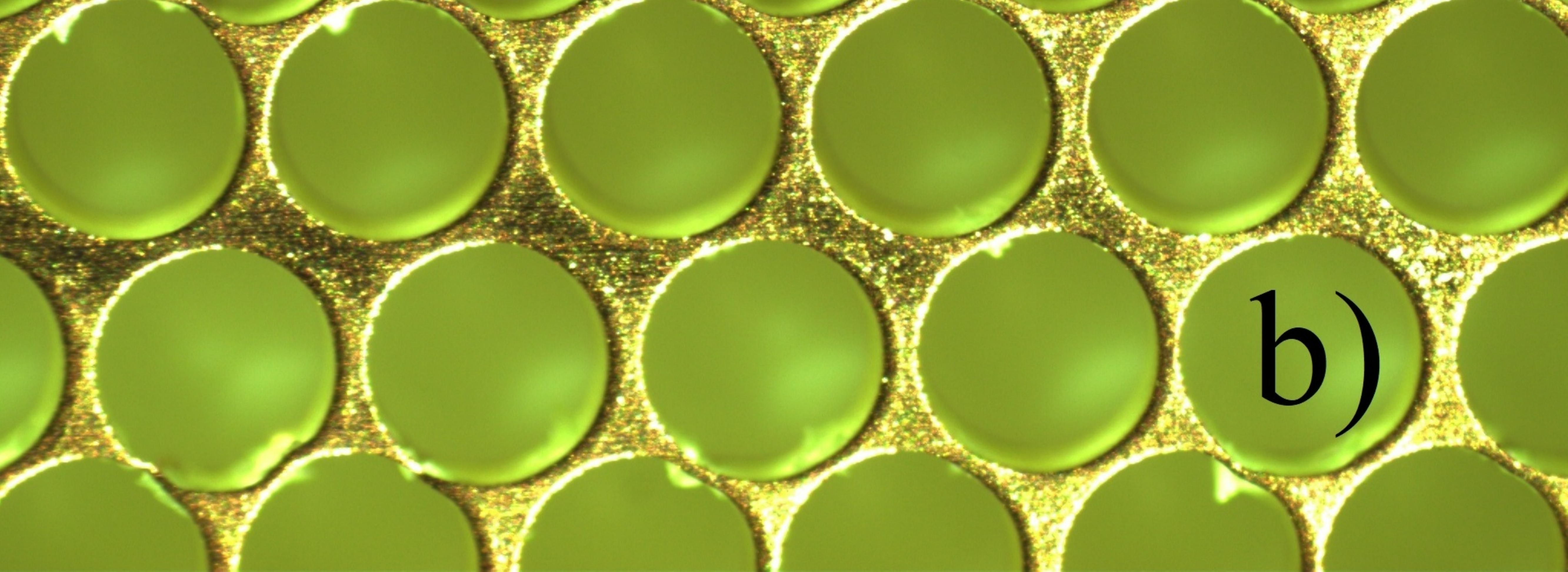}
		\end{tabular}	
	\caption{Images of two types of THGEMs simulated in the present work, taken by a microscope. a) shows THGEM with 28\% optical transparency and b) with 75\% one (refer to Table~\ref{tab:THGEM_pars}).}
	\label{fig:THGEM_photos}
	\end{figure}

Two types of THGEMs were used for calculation of electron transmission with optical transparency of 28\% and 75\% (further referred to as 28\% THGEM and 75\% THGEM): their design parameters are presented in Table \ref{tab:THGEM_pars}. Fig. \ref{fig:THGEM_photos} shows the images of these two types of THGEMs obtained with a microscope. At the moment, the 28\% THGEMs are installed in the detector.

	\begin{table}[t!]
		\begin{center}
		\vspace{-10pt}
		\caption{THGEM geometrical parameters used in simulations. 28\% THGEM is used in the current detector version shown in Fig. \ref{fig:setup_scheme}.}
		\label{tab:THGEM_pars}
			\begin{tabular}{l|c|c}
			 THGEM type & 28\%THGEM & 75\%THGEM \\
			\hline
			Hole pitch [\(\mu\)m] & 900 & 1100 \\
			Hole diameter [\(\mu\)m] & 500 & 1000 \\
			Hole rim [\(\mu\)m] & 100 & 0 \\
			Dielectric  & 400 & 400 \\
			thickness [\(\mu\)m] & & \\
			Copper layer  & 30 & 30 \\
			thickness [\(\mu\)m] & & \\
			Optical & 28\% & 75\% \\
			transparency & & \\
			Manufacturer & CERN & Electroconnect \\
			& & (Russia) \\
			\hline
			\end{tabular}
		\end{center}
	\vspace{-20pt}
	\end{table}

\begin{figure}[!t]
	\includegraphics[width=\linewidth]{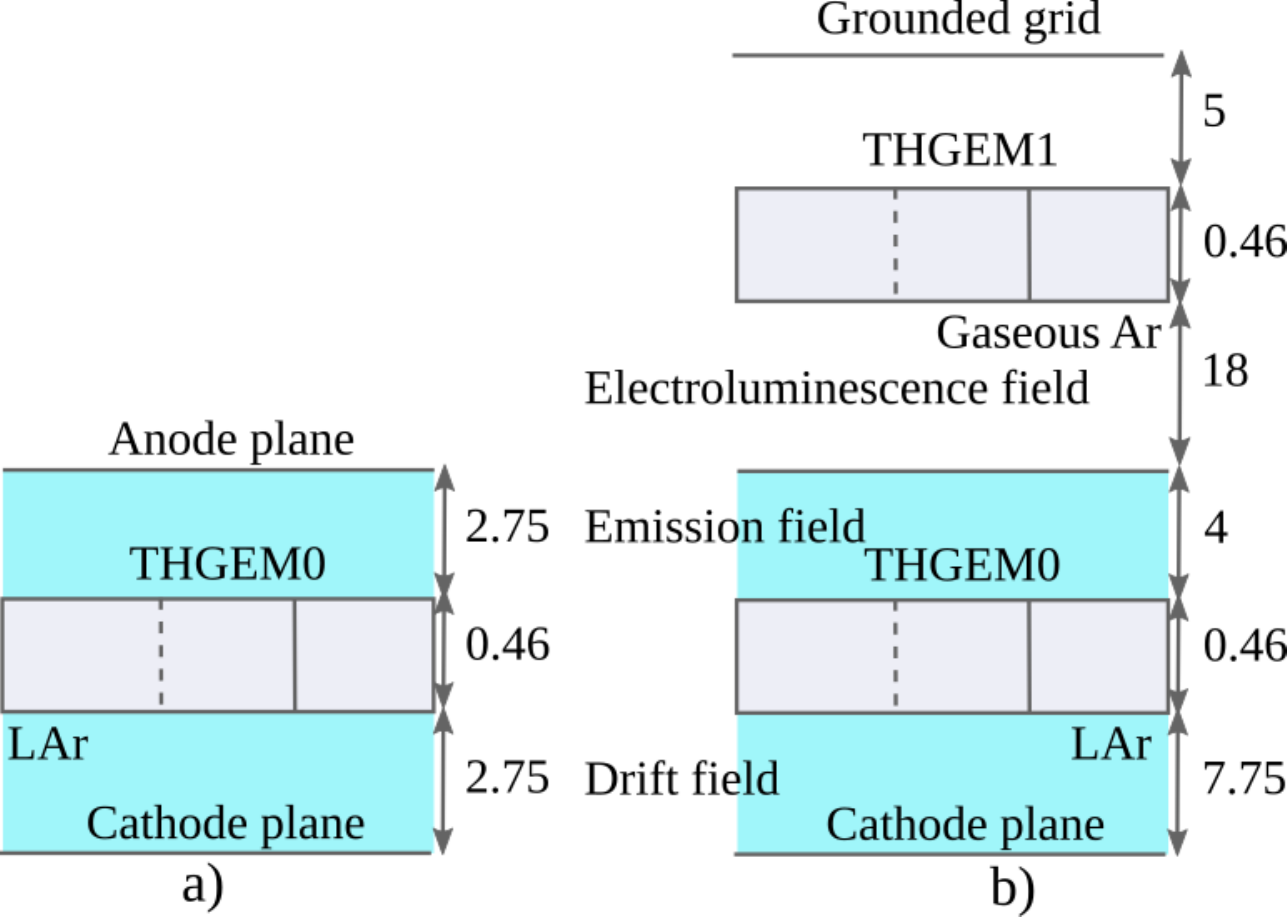}
	\caption{Schematics of simulation geometry for single (a) and double (b) THGEM configuration (not to scale). All sizes are given in mm.}
	\label{fig:model_scheme}
	\vspace{-10pt}
	\end{figure}

The schematics of simulation geometry to determine the electron transmission through THGEM0 in a single-THGEM configuration is presented in Fig. \ref{fig:model_scheme}(a). It should be mentioned that distances from THGEM to the anode and cathode planes are smaller than in the real detector. This is done in order to decrease the size of geometrical model and because there is no interest in calculating electric field at large distances where it becomes uniform. Hence, the voltages at the anode and cathode planes in single-THGEM configuration are not defined by the voltage divider shown in Fig. \ref{fig:setup_scheme} directly but instead are set to provide the same drift and emission fields as in the detector. Since the electron transmission is defined mostly by the voltage across THGEM0, its optimization was conducted by varying the $R_{THGEM0}$ resistance. The $R_{THGEM0}$ value is rather small compared to that of the whole divider and thus the drift, electron emission and EL fields almost do not change during such variations. 
	
In a double-THGEM configuration (see Fig. \ref{fig:model_scheme}(b)), only 28\% THGEMs were used in simulations, along with the standard voltage divider of Fig. \ref{fig:setup_scheme}. As in the single-THGEM case, cathode plane was moved closer to THGEM0 while the rest of parameters correspond to Fig. \ref{fig:setup_scheme}. In these simulations, the voltages applied across THGEM1 were those used in experiment for THGEM/SiPM-matrix readout \cite{SiPMMatrix19}: 2200, 2000 and 0 V. The first two cases correspond to THGEM1 operation in electron avalanche mode, while the latter one is equivalent to the configuration when THGEM1 acted as an anode of the EL gap. 
 
For simulation of electric fields in the whole volume of the detector, THGEMs were substituted by solid dielectric plates with conducting surfaces corresponding to the THGEM electrodes' active area of 10$\times$10 cm$^2$. Such elements as voltage divider and its wires were omitted from the geometry. For simplicity, only elements which strongly influence electric field in the drift, emission and EL regions were included in the simulation.

  \section{Results}

  \subsection{Single-THGEM configuration}

\begin{figure}[t!]
	\includegraphics[width=\linewidth]{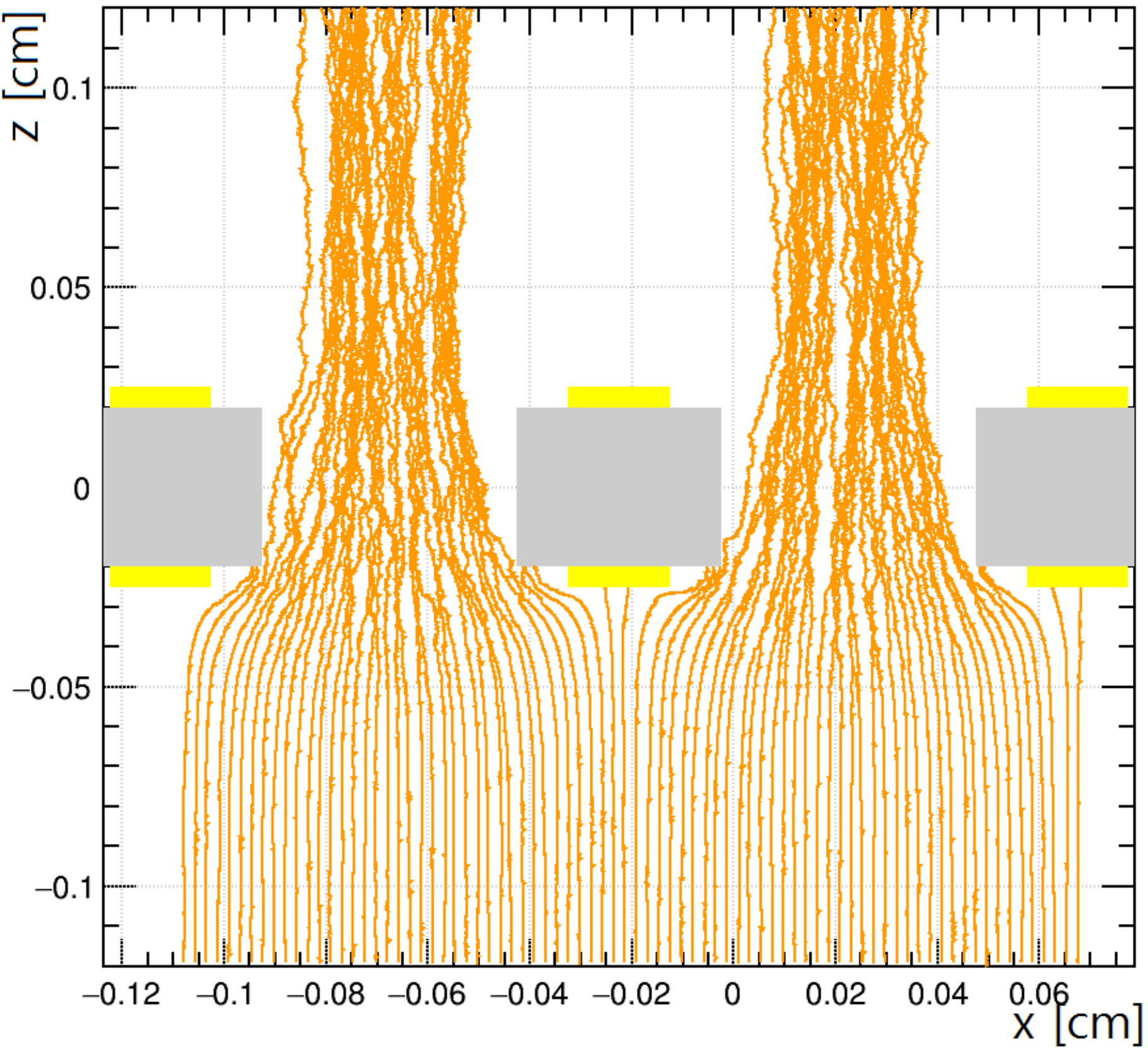}
	\caption{Electron drift trajectories in 28\% THGEM0 in liquid Ar. \(E_{EL} =\) 7.3 kV/cm, \(E_{drift} =\) 0.62 kV/cm, \(E_{emission} =\) 4.8 kV/cm, \(E_{THGEM0} =\) 2.48 kV/cm. Corresponding electron transmission is 61.2\%. Electron diffusion is turned on. The data on diffusion coefficients and drift velocity are discussed in \ref{ToolsAndParameters}.}
	\label{fig:drift_28percent}
	\vspace{-10pt}
	\end{figure}

An example of electron drift trajectories through 28\% THGEM0 with diffusion turned on is presented in Fig.~\ref{fig:drift_28percent}. The simulation uncertainties of $T_e$ were estimated by comparing the results with electron diffusion turned on and off since there are significant uncertainties in diffusion coefficients, see section \ref{ToolsAndParameters}. Every configuration was simulated using 10000 electrons which corresponds to the absolute statistical error of 0.5\% for $T_e$ = 50\% and 0.1\% for $T_e$ = 99\%. As can be seen, statistical error is significantly lower than systematical one.

\begin{figure}[t!]
	\includegraphics[width=\linewidth]{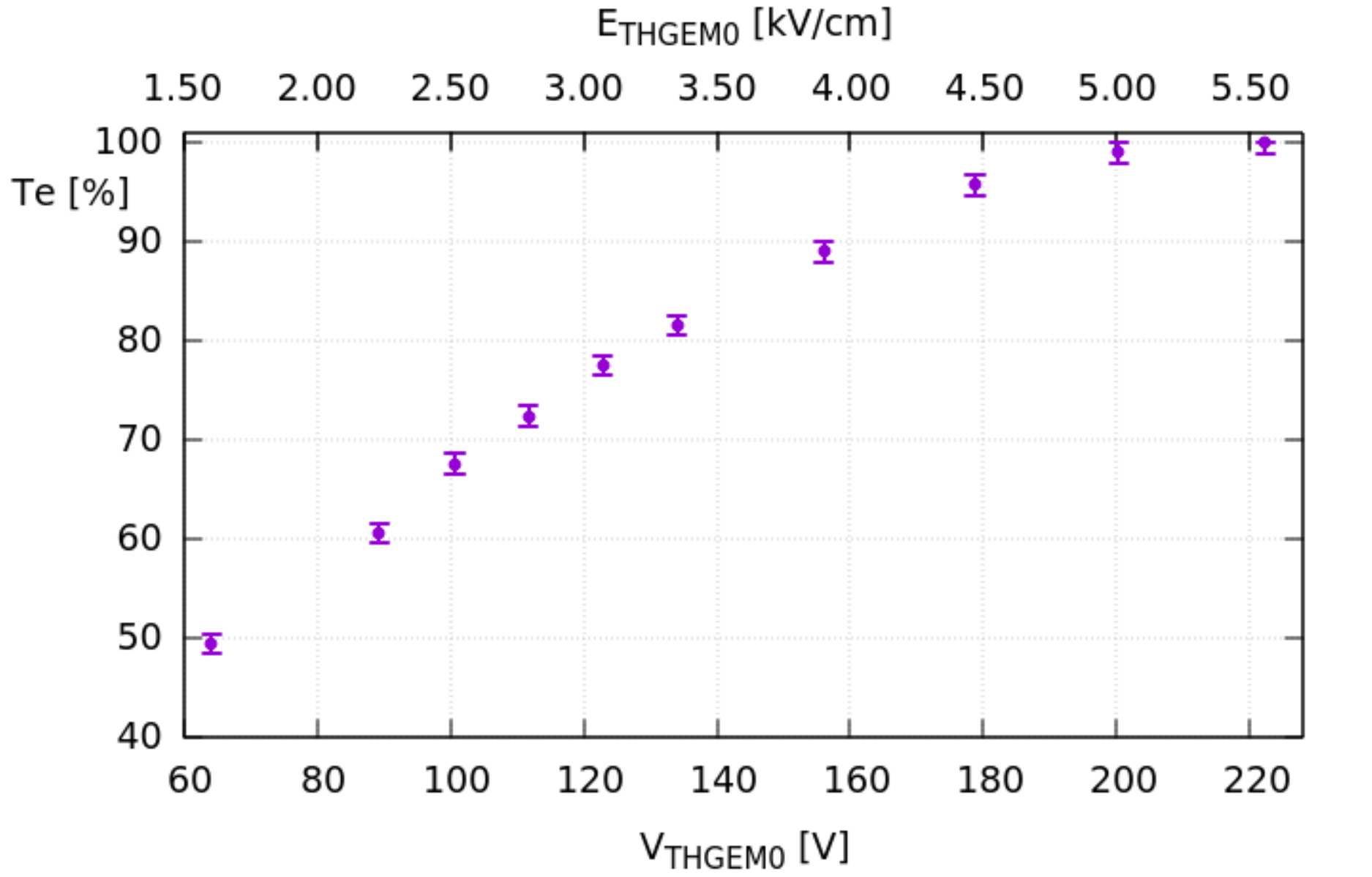}
	\caption{Electron transmission through 28\% THGEM0 in liquid Ar as a function of the voltage across THGEM0 with \(V_{0}\) fixed at 18 kV. The electric fields below and above THGEM0 are practically fixed: \(E_{drift} \approx\) 0.56 kV/cm, \(E_{emission} \approx\) 4.3 kV/cm and \(E_{EL} \approx\) 6.6 kV/cm. Corresponding $R_{THGEM0}$ varies from 3 to 10 M\(\Omega\).}
	\label{fig:Tr_over_V_28}
	\end{figure}

Fig.~\ref{fig:Tr_over_V_28} illustrates how electron transmission depends on the THGEM0 voltage, i.e. on varying $R_{THGEM0}$ in the voltage divider, at fixed \(V_{0}\), the electric fields below and above the THGEM0 being practically fixed. The influence of varying $R_{THGEM0}$ from 3 to 10 M\(\Omega\) on drift and emission fields is negligible because 10\% variation of either of the fields results in the absolute variation of electron transmission less than 0.5\%. One can see that the 61\% electron transmission through THGEM0 with unmodified divider can be easily increased to a maximum of 100\% just by increasing the $R_{THGEM0}$ value from 4 to 10 M\(\Omega\), i.e. changing $E_{drift}$:\hspace{0pt}$E_{THGEM0}$:\hspace{0pt}$E_{emission}$ ratio from 1.0:4.0:7.6 to 1.0:10.0:7.6

\begin{figure}[t!]
	\includegraphics[width=\linewidth]{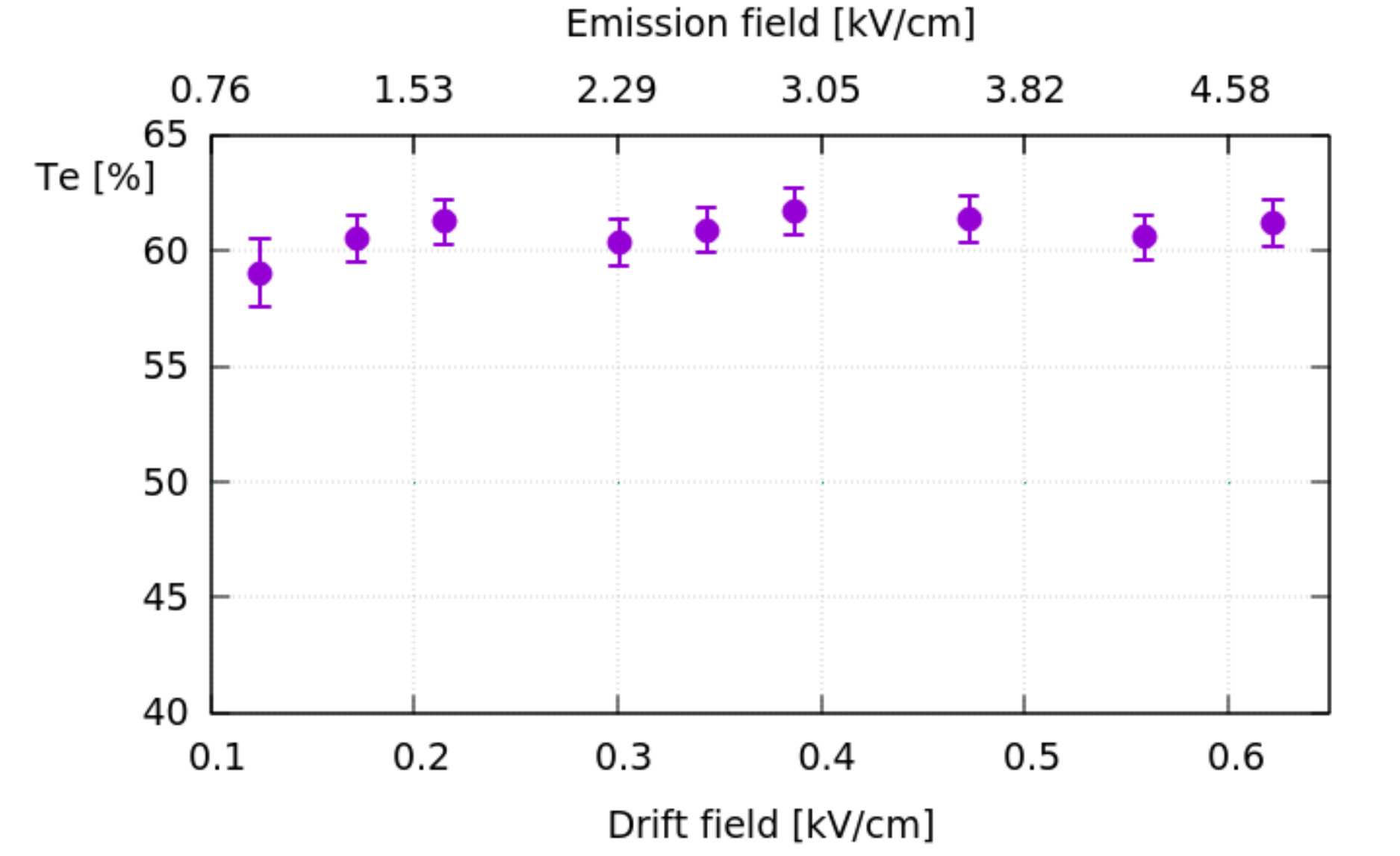}
	\caption{Electron transmission through 28\% THGEM0 in liquid Ar at varying \(V_{0}\) and fixed $R_{THGEM0}$= 4 M\(\Omega\), as a function of the drift field. The corresponding \(E_{emission}\) is shown on the top axis. The ratio $E_{drift}$:\hspace{0pt}$E_{THGEM0}$:\hspace{0pt}$E_{emission}$ is 1.0:4.0:7.6.}
	\label{fig:Tr_over_E_28_4M}
	\end{figure}

The dependence of electron transmission through 28\% THGEM0 in liquid Ar on the drift and emission fields, at varying voltage applied to the voltage divider (\(V_{0}\)), is presented in Fig.~\ref{fig:Tr_over_E_28_4M}. Drift, emission and EL fields, as well as the THGEM0 voltage, vary in proportion to each other according to the voltage divider configuration and geometry depicted in Fig.~\ref{fig:setup_scheme} ($E_{drift}$:\hspace{0pt}$E_{THGEM0}$:\hspace{0pt}$E_{emission}$:\hspace{0pt}$E_{EL}$ = 1.0:4.0:7.6:11.8). As can be seen, the electron transmission remains at 61\%, regardless of $E_{drift}$ changing from 0.12 to 0.62 kV/cm (\(V_{0}\) changing from 3.8 to 20 kV), within the absolute error of 1\%.

	\begin{figure}[t!]
	\includegraphics[width=\linewidth]{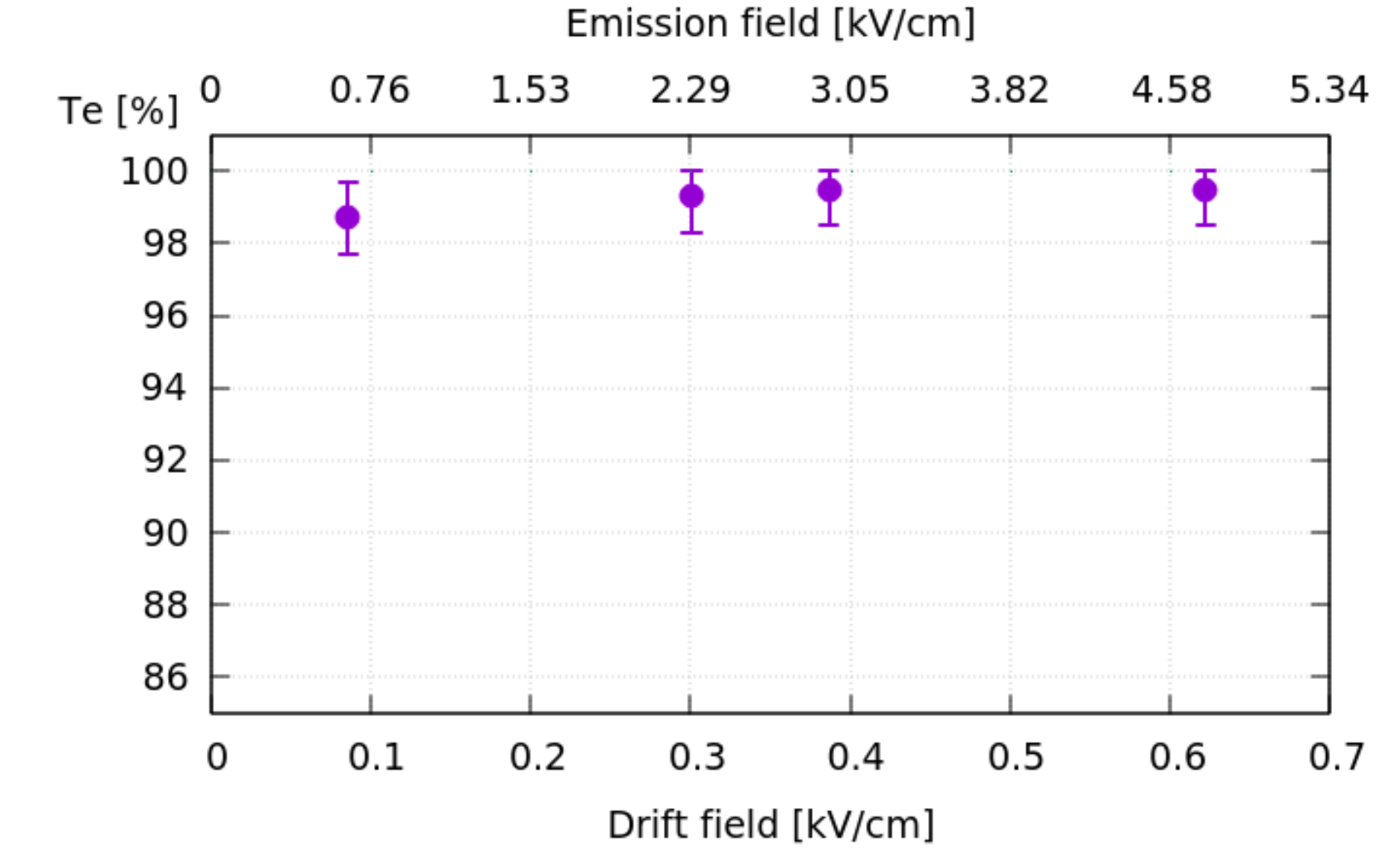}
	\caption{Electron transmission through 75\% THGEM0 in liquid Ar at fixed $R_{THGEM0}$= 4 M\(\Omega\), as a function of the electric field in the EL gap. The corresponding \(E_{emission}\) is shown on the top axis. The ratio $E_{drift}$:\hspace{0pt}$E_{THGEM0}$:\hspace{0pt}$E_{emission}$ is 1.0:4.0:7.6.}
	\label{fig:Tr_over_E_75_4M}
	\vspace{-10pt}
	\end{figure}

Finally, the THGEM0 with high (75\%) optical transparency was studied as a candidate for interface electrode, see Fig.~\ref{fig:Tr_over_E_75_4M}. One can see that even with unmodified voltage divider as in Fig.~\ref{fig:setup_scheme}, i.e. with $R_{THGEM0}$= 4 M\(\Omega\), the electron transmission is close to 100\%, being again independent of \(V_{0}\) and thus of the electric field in the EL gap when $V_{0}$ changes from 3 to 20 kV.

  \subsection{Double-THGEM configuration}

\begin{figure}[t!]
	\includegraphics[width=\linewidth]{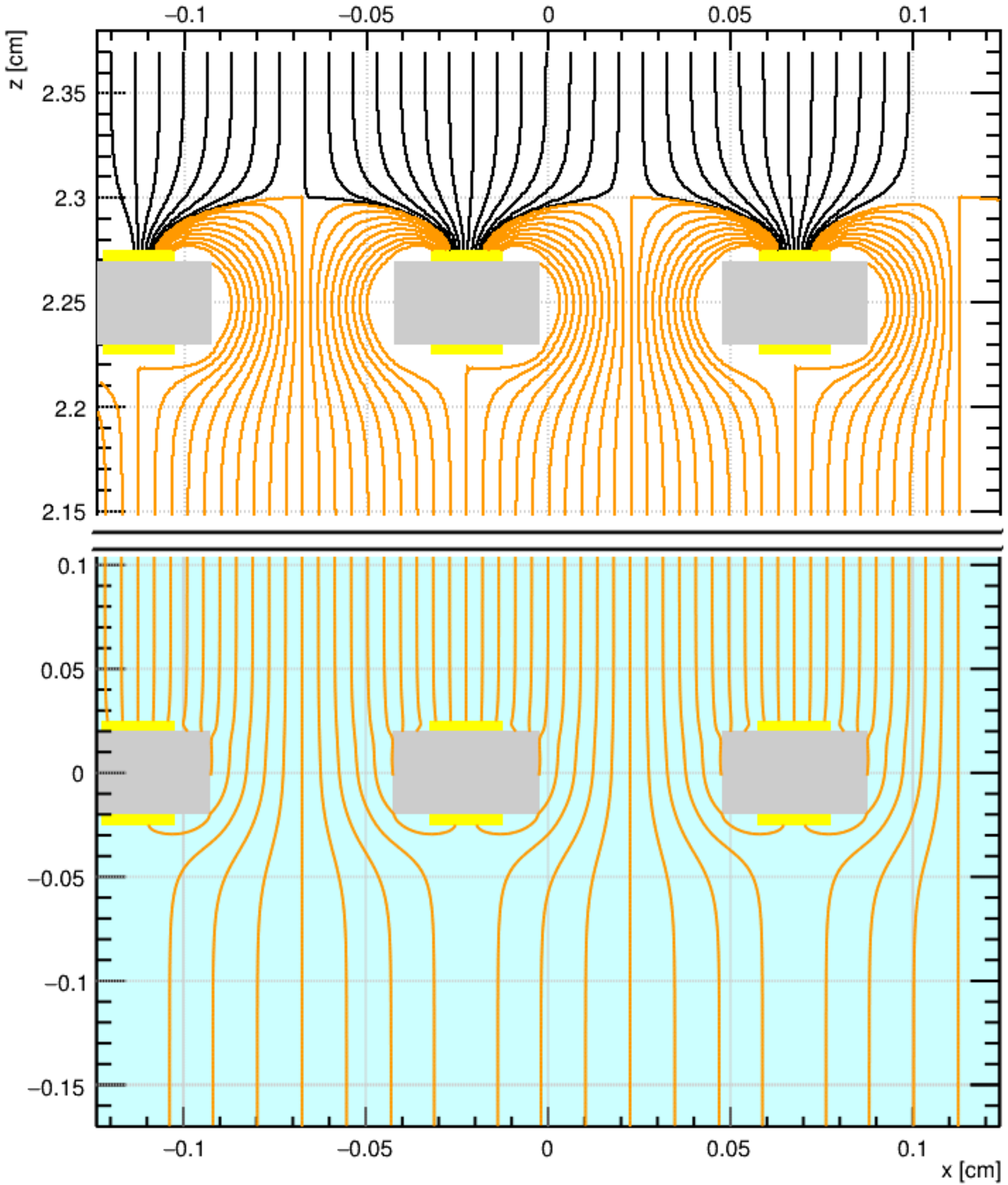}
	\caption{Electric field lines in double-THGEM configuration for \(V_{THGEM1}\) = 2200 V. THGEM0 is in the liquid phase (LAr is depicted by light blue color) and THGEM1 is in the gas one on the top and bottom panel respectively. All field lines start from the topmost copper layer (yellow color). The top plane is the grounded wire grid (see Fig. \ref{fig:setup_scheme}). The area between THGEMs containing gas-liquid interface at 0.42 cm is not shown since it only contains trivial uniform field.}
	\label{fig:double_THGEM_lines_2200}
	\vspace{-10pt}
	\end{figure}

\begin{figure}[t!]
	\includegraphics[width=\linewidth]{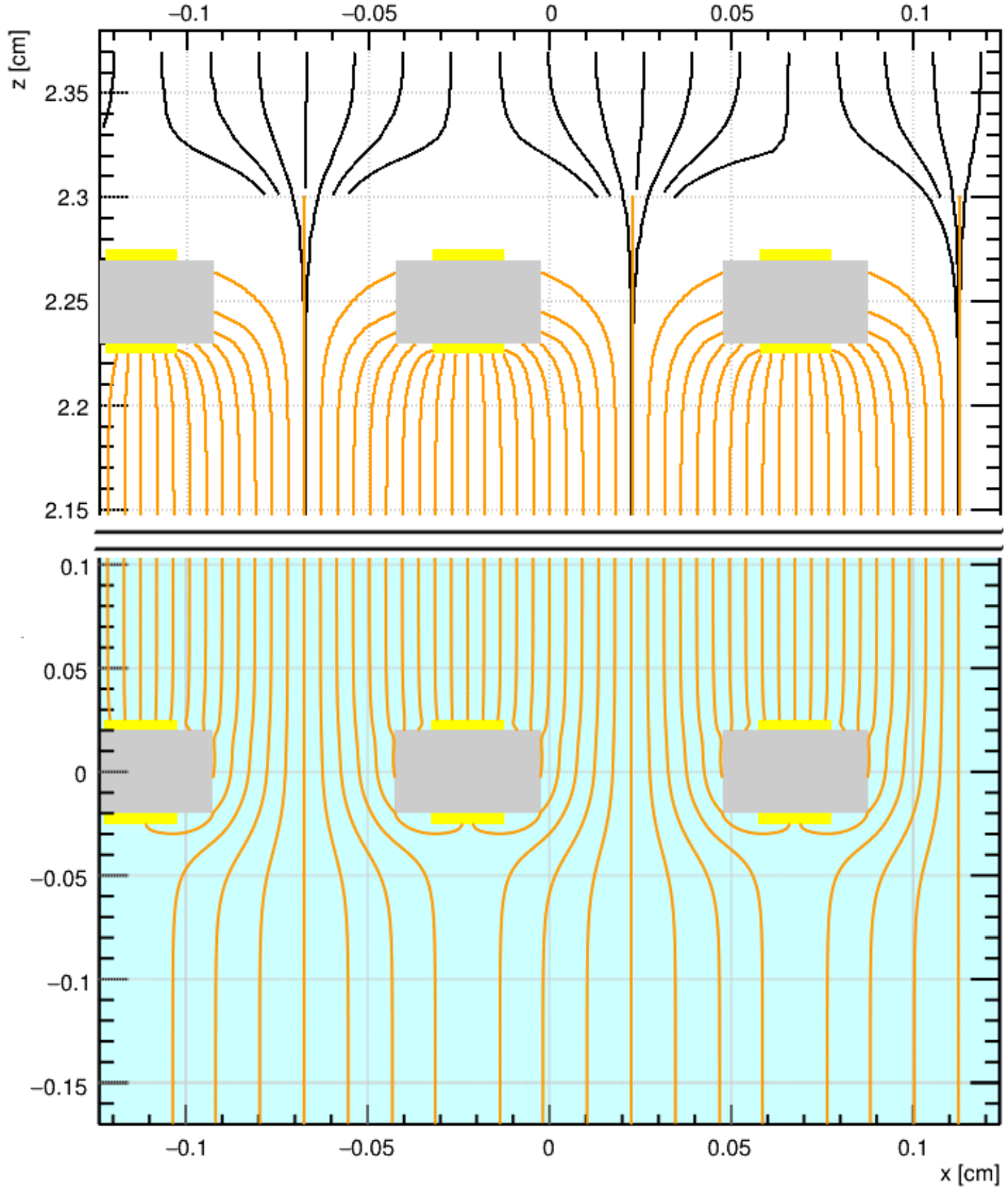}
	\caption{Same as Fig. \ref{fig:double_THGEM_lines_2200} but with \(V_{THGEM1}\) = 0 V.}
	\label{fig:double_THGEM_lines_0}
	\vspace{-10pt}	
	\end{figure}

Electric field lines in double-THGEM configuration are shown in Fig.~\ref{fig:double_THGEM_lines_2200} and Fig.~\ref{fig:double_THGEM_lines_0}. In Fig.~\ref{fig:double_THGEM_lines_2200} the THGEM1, being directly coupled to the EL gap, is operated in electron avalanche mode. In Fig.~\ref{fig:double_THGEM_lines_0} the THGEM1 acts just as an anode of the EL gap. In both cases a typical result for the higher EL field (corresponding to \(V_{0}\)=18 kV) is shown. For other field values, the field pattern (in particular its non-uniformity) is almost the same, because all voltages and fields vary in proportion to each other, according to the given voltage divider configuration. Also, the field patterns for \(V_{THGEM1}\) = 2000 and 2200 V were found to be virtually indistinguishable. 

As can be seen from Fig.~\ref{fig:double_THGEM_lines_2200} and~\ref{fig:double_THGEM_lines_0}, the electric field in the EL gap remains uniform across almost the entire gap, until the distance from the THGEM1 is of the order of its thickness. Let us estimate the effect of the THGEM1 on the field value in the middle of the gap. As expected, the THGEM1 affects it slightly, compared to the values calculated in the infinite plane approximation. In particular, \(E_{EL}\) is 6541, 6535 and 6470 V/cm for \(V_{THGEM1}\) = 2200, 2000 and 0 V respectively, while in the infinite plane approximation  it is 6513 V/cm. Accordingly, this effect is negligible compared to the systematic errors in the absolute EL measurements \cite{NBrS18}.
	
  \subsection{Detector fields}
\begin{figure}[t!]
	\includegraphics[width=\linewidth]{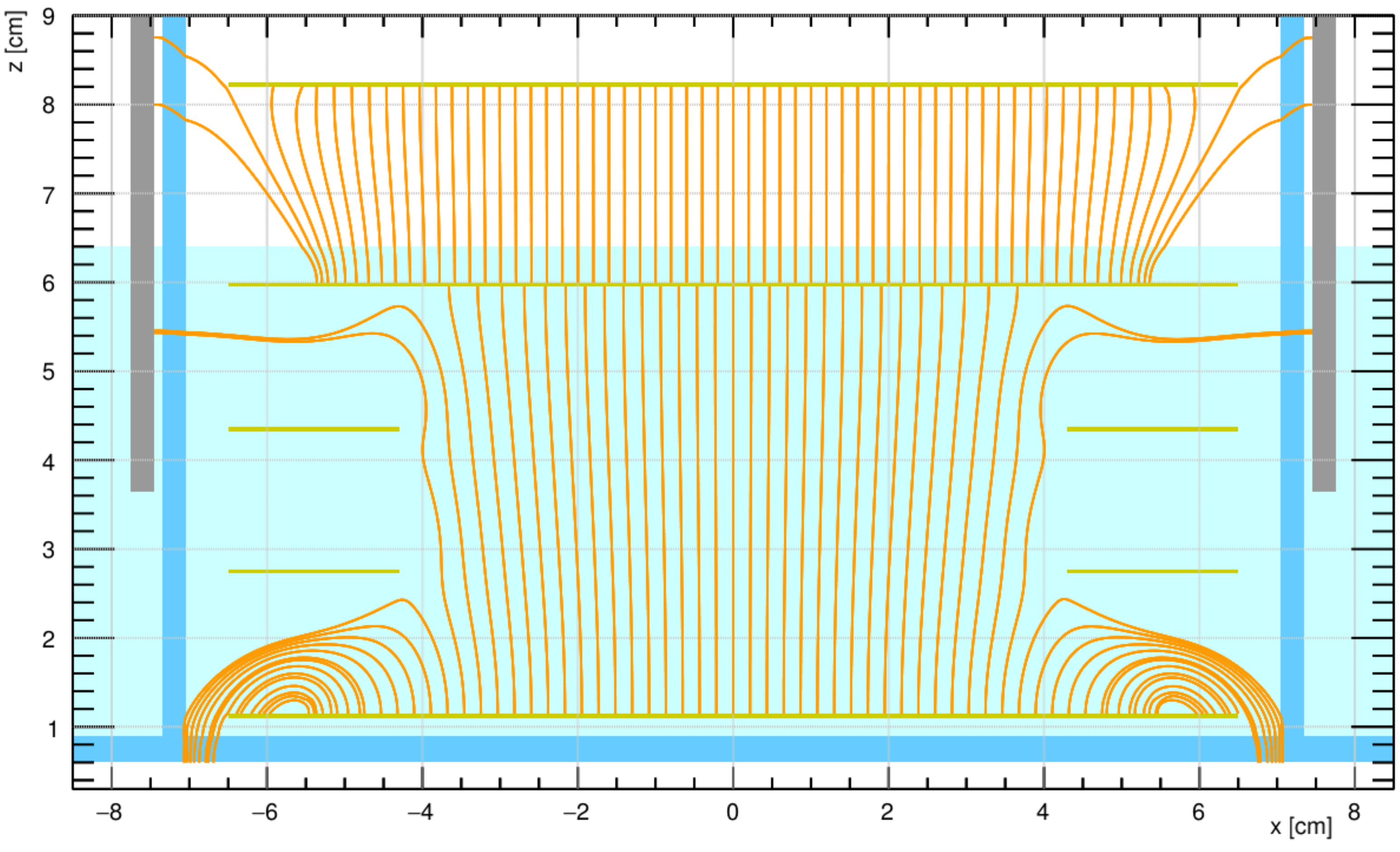}
	\caption{Electric field lines in the volume of the detector for $R_{THGEM0}$= 4 M\(\Omega\) and \(V_{0}\) = 20 kV (to scale). Acrylic is shown in blue color, liquid Ar in light blue and THGEMs and field shaping rings are in dark yellow. Black area denotes grounded metallic box supporting PMTs. Active area of all electrodes is 10$\times$10 cm$^2$.}
	\label{fig:detector_fields_20kV}
	\vspace{-10pt}	
	\end{figure}

The result of simulation of electric fields in the whole drift, electron emission and EL regions is shown in Fig. \ref{fig:detector_fields_20kV}. The simulation was conducted for the configuration with unmodified voltage divider ($R_{THGEM0}$= 4 M\(\Omega\)), \(V_{THGEM1}\) = 0 V and \(V_{0}\) = 20 kV (see Fig. \ref{fig:setup_scheme}). Since all voltages vary in proportion to each other with \(V_{0}\) according to the voltage divider, the field pattern and uniformity are the same for any \(V_{0}\) when \(V_{THGEM1}\) = 0 V.

As Fig. \ref{fig:detector_fields_20kV} demonstrates, the current detector has satisfactory field uniformity in the central area of 50 mm in diameter. In this area, electric field varies by 8\% in the drift region and by 1\% in both emission and EL regions. Hence, the effect of field nonuniformity due to the edge effects of the electrodes is negligible.

  \section{Conclusion}

In this work, the simulations of the electric field of the two-phase detector with electroluminescence (EL) gap have been performed. Also, simulations of electron transport through THGEM electrodes in liquid phase have been done for the first time.

In the liquid phase, these simulations allowed us to determine the optimal parameters, such as the hole diameter of THGEM and applied voltage across it, that can provide the effective transmission of the electrons from the liquid into the EL gap. An important result is that the electron transmission through such an interface THGEM (in the liquid) turned out to be independent of the voltage applied to the voltage divider, i.e. when electric fields change according to ratios $E_{drift}$:\hspace{0pt}$E_{THGEM0}$:\hspace{0pt}$E_{emission}$ = 1.0:4.0:7.6 and 1.0:10.0:7.6, $E_{drift}$ changing from 0.09 to 0.62 kV/cm. Another result is that the 100\% electron transmission can be easily achieved by either an appropriate increase of the resistance of the voltage divider, defining the THGEM voltage, or by using the THGEM with larger hole diameter (i.e. with enhanced optical transparency).

In the gas phase, the effect of the THGEM voltage on the electric field in the EL gap was studied. It was shown that the electric field in the EL gap is uniform nearly everywhere, regardless of the voltage applied across the THGEM, except for distances from the THGEM of the order of its thickness.

Simulations and studies of electric field in the whole detector were also conducted. It was found that edge effects do not affect field uniformity in the active area of the detector.

The results obtained here have been already used in our current studies \cite{NBrS18} and will definitely be used in the future ones.

  \section*{Acknowledgments}

The part of this work regarding simulation of electron transmission through THGEM in liquid Ar was supported by Russian Foundation for Basic Research (project no. 18-02-00117). The rest of the work was supported by Ministry of Science and Higher Education of the Russian Federation. This work was done within the R\&D program of the DarkSide-20k experiment. 

  \bibliography{Manuscript} 
\end{document}